\documentclass[aps,pra,reprint,superscriptaddress]{revtex4-2}
\usepackage{amsmath}
\usepackage{amsthm}
\usepackage{amsfonts}
\usepackage{amssymb}
\usepackage{xcolor,graphicx}
\usepackage{tikz}
\usepackage{color}
\usepackage[pdftex]{hyperref} 
\usepackage{longtable}
\usepackage{array}
\usepackage{dsfont}
\usepackage{braket}
\usepackage{graphicx}
\usepackage{dcolumn}
\usepackage{bm}
\hypersetup{
	colorlinks = true,
	linkcolor = red,
	anchorcolor = red,
	citecolor = red,
	filecolor = red,
	urlcolor = red
}
\begin{document}
\title{Generation of Talbot-like fields}
\author{Jorge A. Anaya-Contreras}
\affiliation{ Instituto Polit´ecnico Nacional, ESFM, Departamento de F´ısica. Edificio 9,
\\ Unidad Profesional Adolfo L´opez Mateos, CP 07738 CDMX, Mexico.}
\author{ Arturo Z\'uñiga-Segundo}
\affiliation{ Instituto Polit´ecnico Nacional, ESFM, Departamento de F´ısica. Edificio 9,
\\ Unidad Profesional Adolfo L´opez Mateos, CP 07738 CDMX, Mexico.}
\author{David S\'anchez-de-la-Llave} 
\affiliation{ Instituto Nacional de Astrofísica Óptica y Electrónica\\Calle Luis Enrique Erro No. 1\\ Santa María Tonantzintla, Pue., 72840, Mexico.}
\author{ H\'ector M. Moya-Cessa}
\affiliation{ Instituto Nacional de Astrofísica Óptica y Electrónica\\Calle Luis Enrique Erro No. 1\\ Santa María Tonantzintla, Pue., 72840, Mexico.}

\begin{abstract}
We present an integral of diffraction based on particular eigenfunctions of the Laplacian in two dimensions.
We show how to propagate some fields, in particular a Bessel field, a superposition of Airy beams, both over the square root of the radial coordinate, and show how to construct a field that reproduces itself periodically in propagation, {\it i.e.}, a field that renders the Talbot effect. Additionally, it is shown that the superposition of Airy beams produces self-focusing.  
\end{abstract}

\date{\today}
\maketitle
\section{Introduction} In recent years there has been much interest in the propagation of light in free space where it has been shown that light not only propagates in straight lines but there are beams that also bend while propagating \cite{berry_1979,PhysRevLett.99.213901,Hwang2010,Cerda2011,Yan2012,Vaveliuk2014,Jauregui2014,Vaveliuk2015,Kaganovsky2010,Papazoglou2015,Torre2015,Aleahmad2016,Weisman2017,Mansour2018,Rozenman2019}.

On the other hand, Montgomery \cite{Montgomery1967} has given the necessary and sufficient conditions  in order that a given field at $z=0$, replicates itself without the aid of lenses or other optical accessories, {\it i.e.}, for the Talbot effect to take place. 

In this contribution, by using eigenfunctions of the perpendicular Laplacian in polar coordinates, we show that an integral of diffraction may be written which we use to propagate some fields, namely Bessel and superposition of Airy beams, both divided by the square root of the radial coordinate. We also show that particular series of Bessel beams with integer or fractional order, reproduce themselves during propagation, {\i.e.}, giving rise to the Talbot effect.

\subsection{Paraxial equation} The paraxial equation reads as
\begin{equation}
    \nabla_{\perp}^2E+2ik \frac{\partial E}{\partial z}=0,
\end{equation}
with solution (where we consider units such that $k = 1$)
\begin{equation}\label{sol}
   E(x,y,z)=\exp\left[ i\frac{z}{2}\nabla_{\perp}^2\right]E(x,y,0)
\end{equation}
where $\nabla_{\perp}$ is the Laplacian that in Cartesian coordinates reads
\begin{equation}
    \nabla_{\perp}^2=\frac{\partial^2}{\partial x^2}+\frac{\partial^2}{\partial y^2}
\end{equation}
and in polar coordinates we may find a set of eigenfunctions
\begin{eqnarray}\label{eigen}
  \left(\frac{\partial^2 }{\partial r^2}+\frac{1}{r}\frac{\partial}{\partial r}+\frac{1}{r^2}\frac{\partial^2}{\partial \theta^2}\right)\frac{\hbox{e}^{\pm i\alpha r}}{\sqrt{r}}e^{\pm i\frac{\theta}{2}}=
  -\alpha^2 \frac{\hbox{e}^{\pm i\alpha r}}{\sqrt{r}}\hbox{e}^{\pm i\frac{\theta}{2}},
\end{eqnarray}
with eigenvalues given by $-\alpha^2$.

\section{Integral of diffraction}
If we consider the field at z = 0 given in the form
\begin{equation}\label{IntOfDiff}
    E(r,\theta,0)=\frac{\hbox{e}^{i\theta/2}}{\sqrt{r}}\int_{-\infty}^{\infty}\mathcal{E}(\alpha)\hbox{e}^{ir\alpha}d\alpha\;,
\end{equation}
it is easy to see that we may write an integral of diffraction of the form
\begin{equation}
    E(r,\theta,z)=\frac{\hbox{e}^{i\theta/2}}{\sqrt{r}}\int_{-\infty}^{\infty}\hbox{e}^{{\color{black}-}i\frac{z}{2}\alpha^2}\mathcal{E}(\alpha)\hbox{e}^{ir\alpha}d\alpha,
\end{equation}
where we have applied the property that a function of (the operator) $\nabla_{\perp}^2$ applied to the eigenfunction is simply the function of the eigenvalue times the eigenfunction, {\it i.e.,}
\begin{eqnarray}\label{0010}
  F(\nabla_{\perp}^2)\frac{\hbox{e}^{i\alpha r}}{\sqrt{r}}\hbox{e}^{\pm i\frac{\theta}{2}}=
  F(-\alpha^2) \frac{\hbox{e}^{i\alpha r}}{\sqrt{r}}\hbox{e}^{\pm i\frac{\theta}{2}}.
\end{eqnarray}

\section{Propagating a Bessel function} We  consider  the following field at $z=0$
\begin{equation}\label{Bess}
   E(r,\theta,z=0)=\frac{J_n(ar)}{\sqrt{r}}\hbox{e}^{i\frac{\theta}{2}}, \qquad n\ge 1,
\end{equation}
where $J_n(x)$ is a Bessel function of order $n$ and the case $n=0$ is not considered because of its singularity. We write the Bessel function in terms of its integral representation
\begin{equation}
   E(r,\theta,z=0)=\frac{1}{2\pi}\int_{-\pi}^{\pi}\frac{\hbox{e}^{iar\sin u}}{\sqrt{r}}\hbox{e}^{i\frac{\theta}{2}}\hbox{e}^{-inu}du,
\end{equation}
such that, by applying the property described by Eq. (\ref{0010}), we obtain
\begin{equation}
   E(r,\theta,z)=\frac{\hbox{e}^{i\frac{\theta}{2}}}{2\pi\sqrt{r}}\int_{-\pi}^{\pi}\hbox{e}^{{\color{black}-}i\frac{a^2z}{2}\sin^2 u}\hbox{e}^{iar\sin u-inu}du.
\end{equation}
It is not difficult to show that the integral above is a so-called Generalized Bessel function. For this, we rewrite it as
\begin{equation}
   E(r,\theta,z)=\frac{\hbox{e}^{{\color{black}-}i\frac{a^2z}{\color{black}4}}\hbox{e}^{i\frac{\theta}{2}}}{2\pi\sqrt{r}}\int_{-\pi}^{\pi}\hbox{e}^{i\frac{a^2z}{4}\cos2 u}\hbox{e}^{iar\sin u-inu}du.
\end{equation}
We define $Z=a^2z/{\color{black}4}$ and use a Taylor series for the cosine term argument exponential, yielding
\begin{equation}
   E(r,\theta,z)=\frac{\hbox{e}^{{\color{black}-}iZ}\hbox{e}^{i\frac{\theta}{2}}}{2\pi\sqrt{r}}\sum_{m=0}^{\infty}\frac{(iZ)^m}{2^mm!}\int_{-\pi}^{\pi}(\hbox{e}^{2iu}+\hbox{e}^{-2iu})^m\hbox{e}^{iar\sin u-inu}du,
\end{equation}
and developing the binomial inside the integral we obtain
\begin{eqnarray}
   E(r,\theta,z)&=&\frac{\hbox{e}^{{\color{black}-}iZ}\hbox{e}^{i\frac{\theta}{2}}}{2\pi\sqrt{r}}\sum_{m=0}^{\infty}\frac{(iZ)^m}{2^mm!}\times\\ \nonumber &&\sum_{k=0}^m
   \left(\begin{array}{c}
        m\\
        k 
   \end{array}\right)\int_{-\pi}^{\pi}\hbox{e}^{2iu(m-2k)}\hbox{e}^{iar\sin u-inu}du.
\end{eqnarray}
We extend the second sum to infinity as we would only add zeros to the sum {\color{black} and} exchange the order of the sums
\begin{eqnarray}\nonumber
   E(r,\theta,z)&=&\frac{\hbox{e}^{{\color{black}-}iZ}\hbox{e}^{i\frac{\theta}{2}}}{2\pi\sqrt{r}}\sum_{k=0}^{\infty}\sum_{m=0}^{\infty}\frac{(iZ)^m}{2^mk!(m-k)!}
  \times\\ && \int_{-\pi}^{\pi}\hbox{e}^{2iu(m-2k)}\hbox{e}^{iar\sin u-inu}du,
\end{eqnarray}
and start the sum that runs on $m$ at $m=k$ (as for $m<k$ the terms added are zero)
\begin{eqnarray}\nonumber
   E(r,\theta,z)&=&\frac{\hbox{e}^{{\color{black}-}iZ}\hbox{e}^{i\frac{\theta}{2}}}{2\pi\sqrt{r}}\sum_{k=0}^{\infty}\sum_{m=k}^{\infty}\frac{(iZ)^m}{2^mk!(m-k)!}
    \times\\ && \int_{-\pi}^{\pi}\hbox{e}^{2iu(m-2k)}\hbox{e}^{iar\sin u-inu}du.
\end{eqnarray}
By letting $j=m-k$ we obtain
\begin{eqnarray}\nonumber
   E(r,\theta,z)&=&\frac{\hbox{e}^{{\color{black}-}iZ}\hbox{e}^{i\frac{\theta}{2}}}{2\pi\sqrt{r}}\sum_{k=0}^{\infty}\sum_{j=0}^{\infty}\frac{(iZ)^{j+k}}{2^{j+k}k!j!}\\
   &\times&\int_{-\pi}^{\pi}\hbox{e}^{2iu(j-k)}\hbox{e}^{iar\sin u-inu}du,
\end{eqnarray}
that, by using the integral representation of Bessel functions, gives
\begin{equation}
   E(r,\theta,z)=\frac{\hbox{e}^{{\color{black}-}iZ}\hbox{e}^{i\frac{\theta}{2}}}{\sqrt{r}}\sum_{k=0}^{\infty}\sum_{j=0}^{\infty}\frac{(iZ)^{j+k}}{2^{j+k}k!j!}
  J_{n+2(k-j)}(ar).
\end{equation}

\begin{figure}
	\begin{center}
		\includegraphics[width=.4\textwidth]{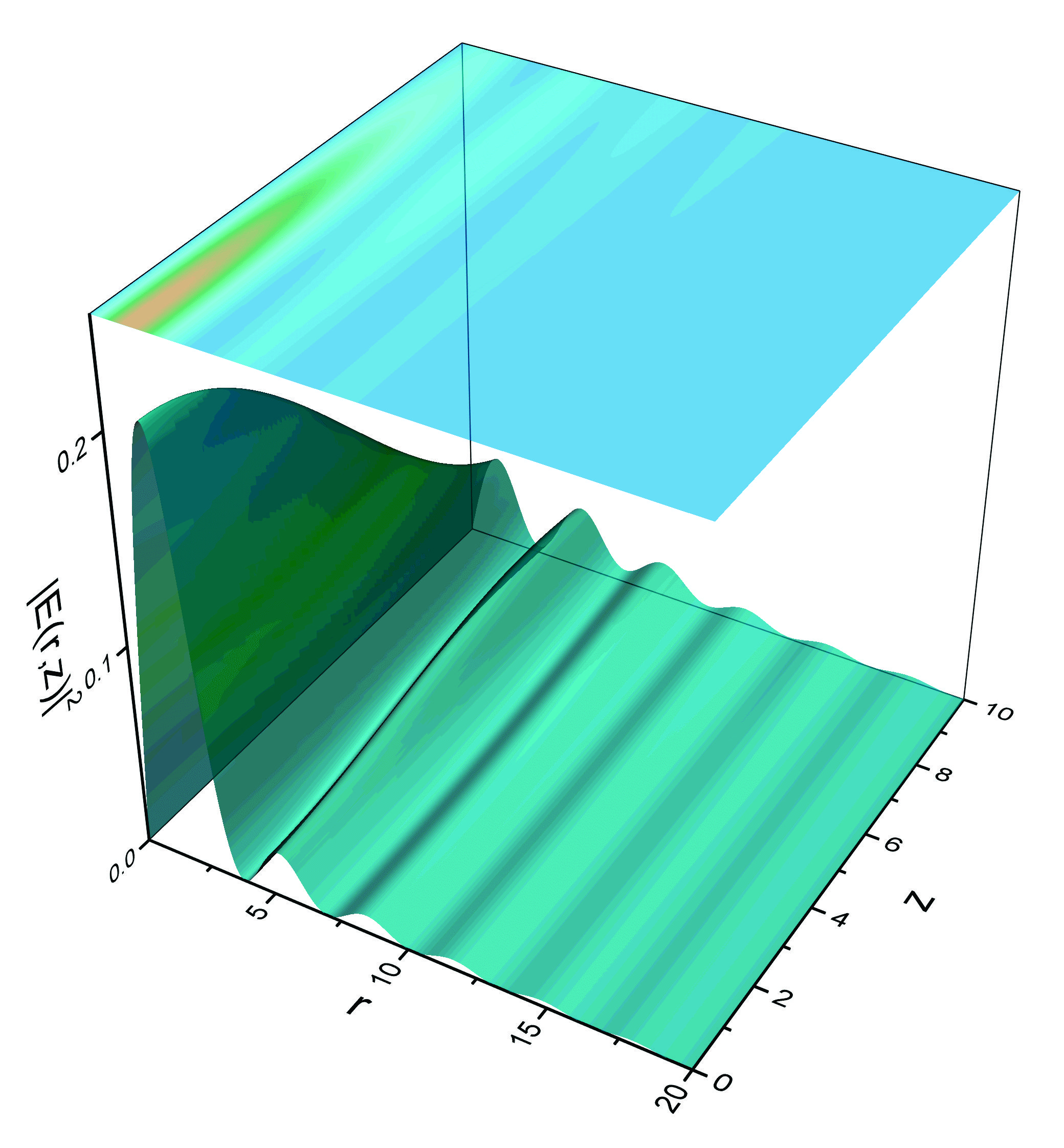} 
		\caption{Intensity field distribution $\vert E(r,z)\vert^2$ obtained from initial state given in Eq. (\ref{Bess}) with $n=1$ and $a=1$.} 
		\label{fig_1}
	\end{center}
\end{figure}
By letting $s=k-j$ we obtain
\begin{equation}
   E(r,\theta,z)=\frac{\hbox{e}^{{\color{black}-}iZ}\hbox{e}^{i\frac{\theta}{2}}}{\sqrt{r}}\sum_{s=-\infty}^{\infty}J_{n+2s}(ar)\sum_{j=0}^{\infty}\frac{(iZ)^{2j+s}}{2^{2j+s}(j+s)!j!},
 \end{equation}
 where we have extended the sum on $s$ to minus infinity as we simply add zeros.

This finally gives a sum of two Bessel functions of different order 
\begin{equation} \label{genebessel}
   E(r,\theta,z)=\frac{\hbox{e}^{{\color{black}-}iZ}\hbox{e}^{i\frac{\theta}{2}}}{\sqrt{r}}\sum_{s=-\infty}^{\infty}i^sJ_{n+2s}(ar)J_s({\color{black}Z}),
 \end{equation}
that may be added to give the so-called Generalized
Bessel functions studied by Dattoli {\it et al.} \cite{Dattoli1989,Dattoli1991} and Eichelkraut \cite{Toni2014}. By using that Bessel generalized functions, given by the expression $\mathcal{J}_n(r,z;g)=\sum_{s=-\infty}^{\infty}g^sJ_{n-2s}(r)J_s(z)$, we write the propagated field as
\begin{equation}
   E(r,\theta,z)=\frac{\hbox{e}^{{\color{black}-}iZ}\hbox{e}^{i\frac{\theta}{2}}}{\sqrt{r}}\mathcal{J}_n(ar,{\color{black}Z};i).
 \end{equation}
 
 In Fig. \ref{fig_1} we plot the field intensity for an initial Bessel function of order $n=1$ as a function of the propagation distance and the radial coordinate. It may be observed that there is an energy redistribution from the central rings towards the outer rings as the field propagates, nevertheless, an overall intensity decrease also exists. 
\begin{figure}
	\begin{center}
	\includegraphics[width=.4\textwidth]{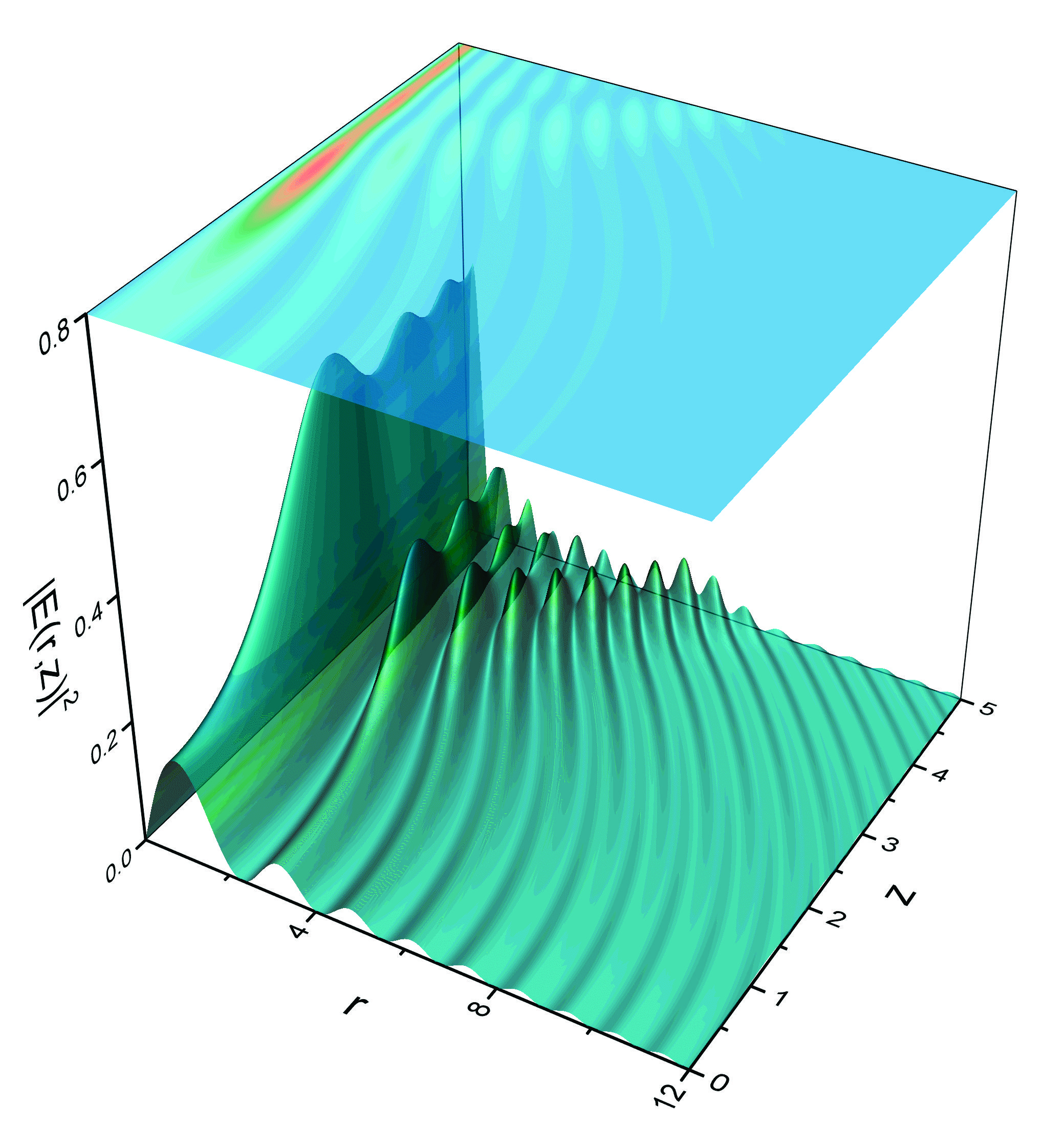} 
		\caption{Intensity field distribution $\vert E(r,z)\vert^2$ obtained from initial state given in Eq.  (\ref{Airy}) .} 
		\label{fig_2}
	\end{center}
\end{figure}

\section{Superposition of Airy functions} We now study the propagation of a superposition of Airy beams \cite{PhysRevLett.99.213901,Vaveliuk2014} whose distribution at $z=0$ is given by
\begin{eqnarray}\label{Airy}\nonumber
 E(r,\theta,z=0)&=&\frac{\hbox{e}^{i\frac{\theta}{2}}}{2\pi\sqrt{r}}\left[\int_{-\infty}^{\infty}\hbox{e}^{i\left(\frac{t^{3}}{3}+rt\right)}\right. dt\\ &-& \left.\int_{-\infty}^{\infty}\hbox{e}^{i\left(\frac{t^{3}}{3}-rt\right)}dt\right],
\end{eqnarray}
where we have written the Airy function in its integral representation. By applying the integral of diffraction given by Eq. (\ref{IntOfDiff}) we obtain
\begin{eqnarray}\nonumber
 E(r,\theta,z)&=&\frac{\hbox{e}^{i\frac{\theta}{2}}}{2\pi\sqrt{r}}\left[\int_{-\infty}^{\infty}\hbox{e}^{i\left(\frac{t^{3}}{3}+rt\right)}\hbox{e}^{{\color{black}-}i\frac{zt^2}{2}}dt\right.\\ &-&\left.\int_{-\infty}^{\infty}\hbox{e}^{i\left(\frac{t^{3}}{3}-rt\right)}\hbox{e}^{{\color{black}-}i\frac{zt^2}{2}}dt\right].
\end{eqnarray}

\begin{figure}
	\begin{center}
	\includegraphics[width=.4\textwidth]{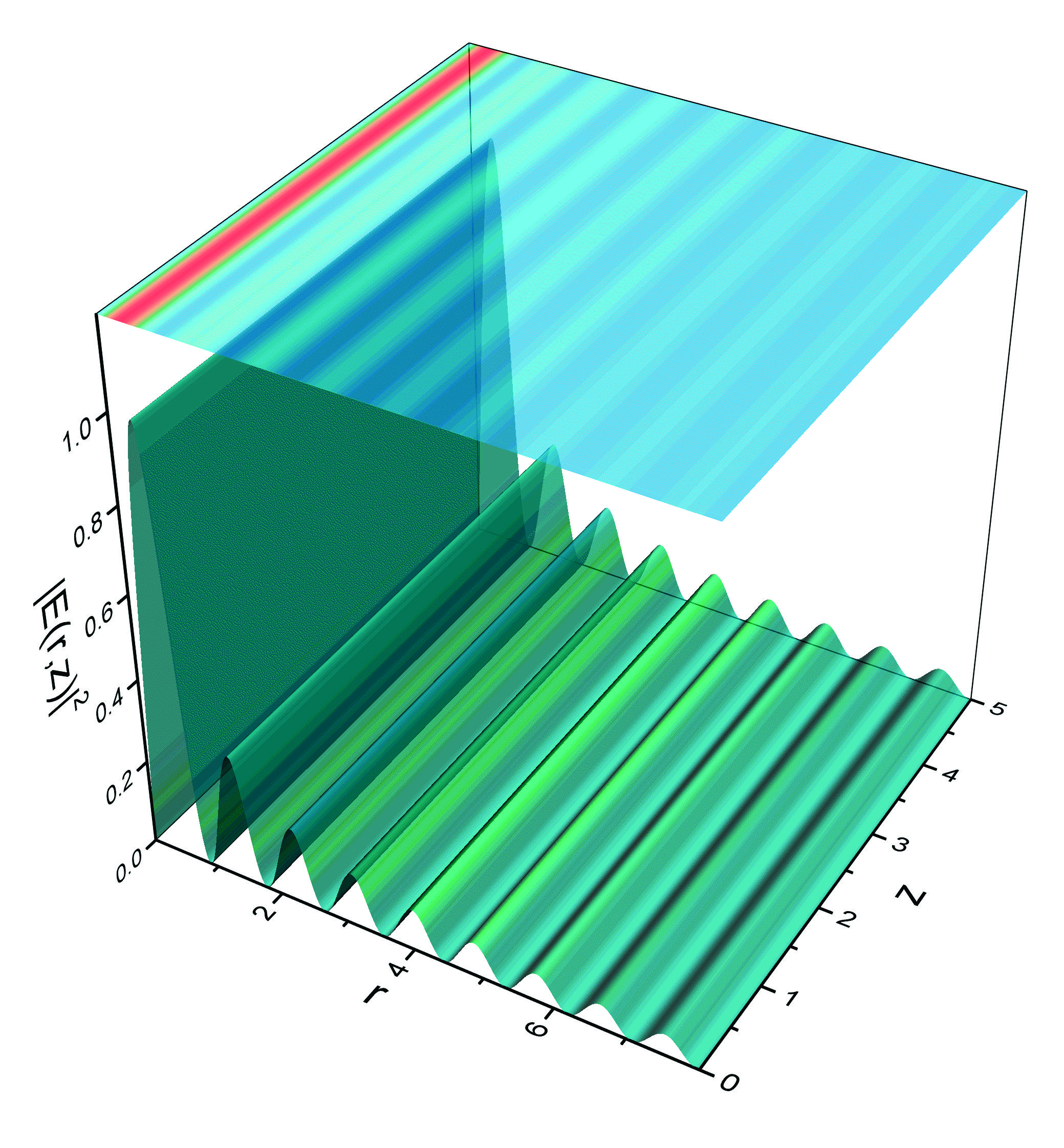} 
		\caption{Normalized Field intensity distribution $\vert E(r,z)\vert^2$ obtained from initial state (\ref{27}).} 
		\label{fig_2.5}
	\end{center}
\end{figure}

By changing variables in the integrals above, we may rewrite them as
\begin{eqnarray}\nonumber
 E(r,\theta,z)&=&\frac{\hbox{e}^{{\color{black}-}i\frac{z^3}{12}}\hbox{e}^{i\frac{\theta}{2}}}{2\pi\sqrt{r}}\left[\hbox{e}^{i\frac{rz}{2}}\int_{-\infty}^{\infty}\hbox{e}^{i\left(\frac{t^{3}}{3}+\left[r{\color{black}-}\frac{z^2}{4}\right]t\right)}dt\right.\\ &-&\left.\hbox{e}^{{\color{black}-}i\frac{rz}{2}}\int_{-\infty}^{\infty}\hbox{e}^{i\left(\frac{t^{3}}{3}-\left[r{\color{black}+}\frac{z^2}{4}\right]t\right)}dt\right],
\end{eqnarray}
that finally yields the superposition of Airy functions
\begin{eqnarray}\nonumber
 E(r,\theta,z)&=&\frac{\hbox{e}^{{\color{black}-}i\frac{z^3}{12}}\hbox{e}^{i\frac{\theta}{2}}}{\sqrt{r}}\left(\hbox{e}^{{\color{black}+}i\frac{rz}{2}}
 \hbox{Ai}\left[r{\color{black}-}\frac{z^2}{4}\right]\right.\\ &-&\left. \hbox{e}^{{\color{black}-}i\frac{rz}{2}}\hbox{Ai}\left[-r{\color{black}-}\frac{z^2}{4}\right]\right).
 \end{eqnarray}
 We plot the propagated field intensity in Fig. \ref{fig_2} where it may be observed  abrupt focusing that may be attributed to the superposition of the Airy functions. There is one Airy function whose main contribution would be in the negative part of the axis, and would bend towards the right. However, as $r$ is always positive, it does not have enough weight to produce an effect. On the other hand, the Airy function whose main contribution is on the positive part, dominates the propagation and bends towards the left. Although there is no medium, the focusing may be explained by the fact that the Airy function {\it produces} an effective index of refraction (the so-called Bohm potential in quantum mechanics) \cite{Asenjo2021,Hojman2021} that gives rise to such behaviour.

\section{Talbot effect}
We can superimpose the eigenfunctions described by Eq. (\ref{eigen}), with the same eigenvalue to find another eigenfunction, a beam of the form $\frac{\sin{br}}{\sqrt{r}}$, which takes us to a Bessel function of order one half
\begin{equation} \label{27}
   E(r,\theta,z=0)=J_{\frac{1}{2}}(br)\hbox{e}^{i\frac{\theta}{2}},
\end{equation}
that is indeed, a diffraction-free beam and we plot it  in Fig. \ref{fig_2.5}.

We then may show that, a proper superposition of them
\begin{equation}
   E(r,\theta,z=0)=\hbox{e}^{i\frac{\theta}{2}}\sum_{k=1}^N c_kJ_{\frac{1}{2}}\left(\sqrt{2^{k+1}\pi}r\right),
\end{equation}

\begin{figure}
	\begin{center}
		\includegraphics[width=.4\textwidth]{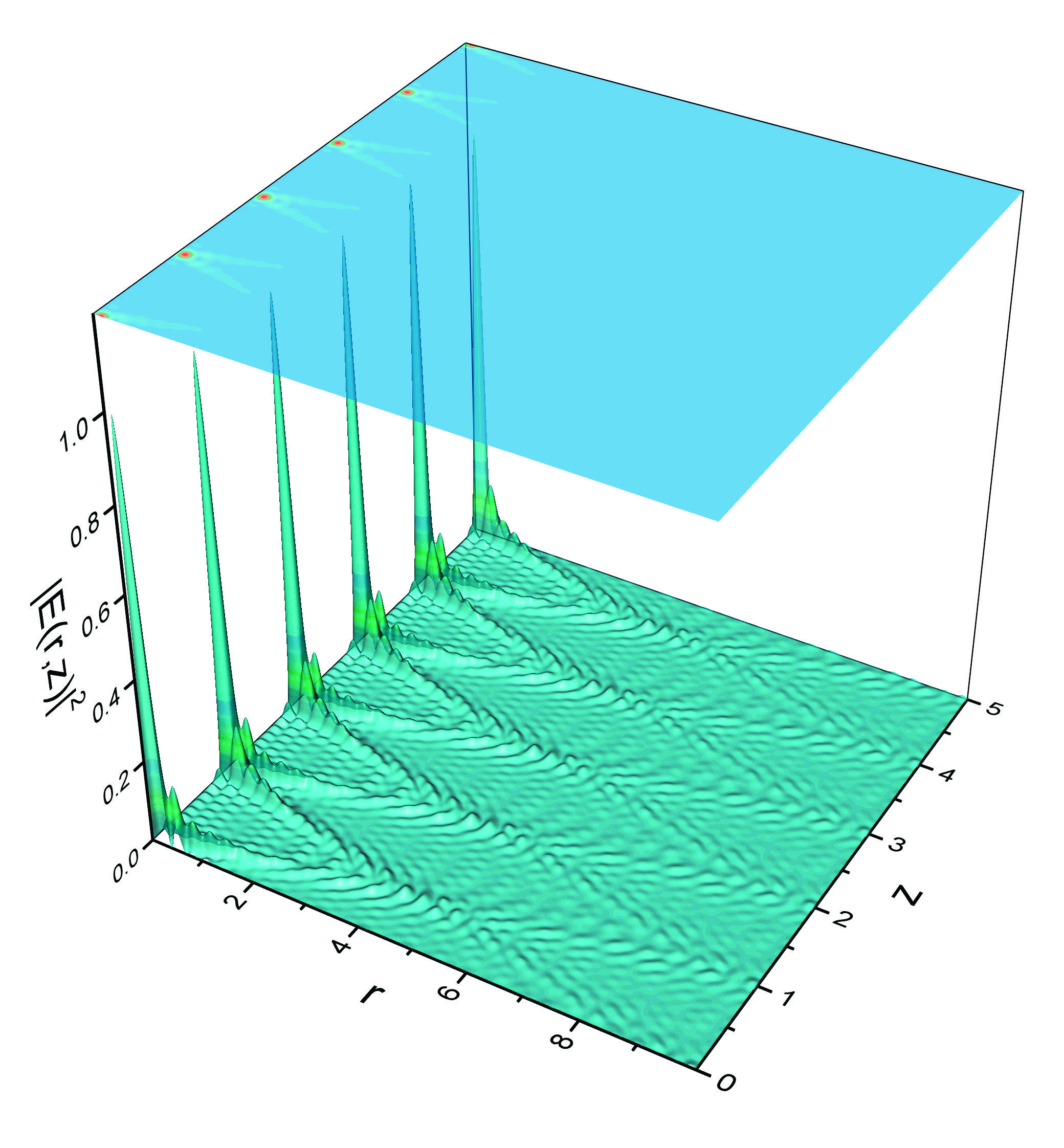} 
		\caption{Normalized Field intensity distribution $\vert E(r,z)\vert^2$ obtained from initial state given by Eq. (\ref{superpo}) with $N=10$ and $c_k=1$.} 
		\label{fig_3}
	\end{center}
\end{figure}
propagates as
\begin{equation}
   E(r,\theta,z)=\hbox{e}^{i\frac{\theta}{2}}\sum_{k=1}^Ne^{-i2^{k}z\pi} c_kJ_{\frac{1}{2}}\left(\sqrt{2^{k+1}\pi}r\right),
\end{equation}
that recovers  periodically the field at $z=0$,
\begin{eqnarray}\nonumber
   E(r,\theta,z=n)&=&\hbox{e}^{i\frac{\theta}{2}}\sum_{k=1}^Ne^{-i2^{k}n\pi} c_k J_{\frac{1}{2}}\left(\sqrt{2^{k+1}\pi}r\right)\\&=&E(r,\theta,z=0).
\end{eqnarray}
We plot in Fig. \ref{fig_3} the field propagated for $c_k=1$ and $N=10$, where it may be seen clearly this effect.

\subsection{Generalization of the Talbot effect to any order of the Bessel function} It  is well-known that Bessel functions (of integer or fractional order) obey the differential equation \cite{Gradshtey}
\begin{equation}
    \frac{d^2J_{\nu}(\beta r)}{dr^2}+\frac{1}{r}\frac{dJ_{\nu}(\beta r)}{dr}+\left(\beta^2-\frac{\nu^2}{r^2}\right)J_{\nu}(\beta r)=0,
\end{equation}
which, if multiplied by $\hbox{e}^{i\nu\theta}$, may be rewritten as
\begin{equation}
   \left( \frac{d^2}{dr^2}+\frac{1}{r}\frac{d}{dr}+\frac{1}{r^2}\frac{d^2}{d\theta^2}\right)J_{\nu}(\beta r)\hbox{e}^{i\nu\theta}=-\beta^2J_{\nu}(\beta r)\hbox{e}^{i\nu\theta},
\end{equation}
or
\begin{equation}
   \nabla_{\perp}^2J_{\nu}(\beta r)\hbox{e}^{i\nu\theta}=-\beta^2J_{\nu}(\beta r)\hbox{e}^{i\nu\theta},
\end{equation}
making the functions $J_{\nu}(\beta r)\hbox{e}^{i\nu\theta}$ eigenfunctions (with eigenvalue $-\beta^2$) of the Laplacian in polar coordinates and therefore becoming diffraction-free beams \cite{Durnin1987}. Therefore a field at $z=0$ given by
\begin{equation}\label{superpo}
   E(r,\theta,z=0)=\hbox{e}^{i\nu\theta}\sum_{k=1}^N c_kJ_{\nu}\left(\sqrt{2^{k+1}\pi}r\right),
\end{equation}
propagates as
\begin{equation}
   E(r,\theta,z)=\hbox{e}^{i\nu\theta}\sum_{k=1}^N\hbox{e}^{-i2^{k}z\pi} c_kJ_{\nu}\left(\sqrt{2^{k+1}\pi}r\right),
\end{equation}
recovering Eq. (\ref{27})  for $\nu=1/2$. The field  at $z=0$, is then recovered periodically, {\it i.e.},
\begin{eqnarray}\nonumber
   E(r,\theta,z=n)&=&\hbox{e}^{i\nu\theta}\sum_{k=1}^N\hbox{e}^{-i2^{k}n\pi} c_k J_{\nu}\left(\sqrt{2^{k+1}\pi}r\right)\\ &=& E(r,\theta,z=0),
\end{eqnarray}
as seen in Fig. \ref{fig_3}

\section{Conclusions}  We have shown that by properly writing a field at $z=0$ we may propagate it by using an integral of diffraction that we introduced in this manuscript. We have shown how to propagate Bessel and a superposition of Airy beams (over the square root of the radial coordinate) and have shown that a series of Bessel functions that may have integer or fractional order  and with proper parameters reproduces itself during propagation, therefore producing the Talbot effect. We have shown self focusing of the superposition of Airy beams that may be explained by the existence of an {\it effective} index of refraction related to the Bohm potential. 

\begin{thebibliography}{9}

\bibitem{berry_1979} M. V. Berry and N. L. Balazs, “Nonspreading wave packets,” Am. J.
Phys. 47, 264–267 (1979).
\bibitem{PhysRevLett.99.213901} G. A. Siviloglou, J. Broky, A. Dogariu, and D. N. Christodoulides, “Observation
of accelerating airy beams,” Phys. Rev. Lett. 99, 213901
(2007).
\bibitem{Hwang2010} C.-Y. Hwang, D. Choi, K.-Y. Kim, and B. Lee, “Dual airy beam,” Opt.
Express 18, 23504–23516 (2010).
\bibitem{Cerda2011} S. Chávez-Cerda, U. Ruiz, V. Arrizón, and H. M. Moya-Cessa, “Generation
of airy solitary-like wave beams by acceleration control in inhomogeneous
media,” Opt. Express 19, 16448–16454 (2011).
\bibitem{Yan2012} S. Yan, B. Yao, M. Lei, D. Dan, Y. Yang, and P. Gao, “Virtual source for
an airy beam,” Opt. Lett. 37, 4774–4776 (2012).
\bibitem{Vaveliuk2014} P. Vaveliuk, A. Lencina, J. A. Rodrigo, and O. M. Matos, “Symmetric
airy beams,” Opt. Lett. 39, 2370–2373 (2014).
\bibitem{Jauregui2014} R. Jáuregui and P. Quinto-Su, “On the general properties of symmetric
incomplete airy beams,” J. Opt. Soc. Am. A 31, 2484–2488 (2014).
\bibitem{Vaveliuk2015} P. Vaveliuk, A. Lencina, J. A. Rodrigo, and O. M. Matos, “Intensitysymmetric
airy beams,” J. Opt. Soc. Am. A 32, 443–446 (2015).
\bibitem{Kaganovsky2010} Y. Kaganovsky and E. Heyman, “Wave analysis of airy beams,” Opt.
Express 18, 8440–8452 (2010).
\bibitem{Papazoglou2015} D. G. Papazoglou, V. Y. Fedorov, and S. Tzortzakis, “Janus waves,” Opt.
Lett. 41, 4656–4659 (2016).
\bibitem{Torre2015} A. Torre, “Propagating airy wavelet-related patterns,” J. Opt. 17, 075604
(2015).
\bibitem{Aleahmad2016} P. Aleahmad, H. Moya-Cessa, I. Kaminer, M. Segev, and D. N.
Christodoulides, “Dynamics of accelerating bessel solutions of
maxwell’s equations,” J.Opt. Soc. Am. A 33, 2047–2052 (2016).
\bibitem{Weisman2017} D. Weisman, S. Fu, M. Gonçalves, L. Shemer, J. Zhou, W. P. Schleich,
and A. Arie, “Diffractive focusing of waves in time and in space,” Phys.
Rev. Lett. 118, 154301 (2017).
\bibitem{Mansour2018} D. Mansour and D. G. Papazoglou, “Tailoring the focal region of abruptly
autofocusing and autodefocusing ring-airy beams,” OSA Continuum 1,
104–115 (2018).
\bibitem{Rozenman2019} G. G. Rozenman, M. Zimmermann, M. A. Efremov, W. P. Schleich,
L. Shemer, and A. Arie, “Amplitude and phase of wave packets in a
linear potential,” Phys. Rev. Lett. 122, 124302 (2019).
\bibitem{Montgomery1967} W. D. Montgomery, “Self-imaging objects of infinite aperture,” J.Opt.
Soc. Am. 57, 772–778 (1967).
\bibitem{Dattoli1989} G. Dattoli, L. Giannessi, M. Richetta, and A. Torre, “Miscellaneous
results on infinite series of bessel functions,” Il Nuovo Cimento 103B,
149–159 (1989).
\bibitem{Dattoli1991} G. Dattoli, A. Torre, S. Lorenzutta, G. M. S., and C. Chiccoli, “Theory of
generalized bessel functions.-ii,” Il Nuovo Cimento 101, 21–51 (1991).
\bibitem{Toni2014} T. Eichelkraut, C. Vetter, A. Perez-Leija, H. Moya-Cessa, D. N.
Christodoulides, and A. Szameit, “Coherent random walks in free
space,” Optica 1, 268–271 (2014).

\bibitem{Asenjo2021} F. A.Asenjo, S. A. Hojman, H. M. Moya-Cessa, and F. Soto-Eguibar,
“Propagation of light in linear and quadratic grin media: The bohm
potential,” Opt. Commun. 490, 126947 (2021).
\bibitem{Hojman2021}  Sergio A. Hojman, Felipe A. Asenjo, and H.M. Moya-Cessa and F. Soto-Eguibar,
“Bohm potential is real and its effects are measurable,” Optik 232,
166341 (2021).
\bibitem{Gradshtey} I. Gradshteyn and I. Ryzhik, Table of integrals, series, and products
(Academic Press, Inc., San Diego, 1980).
\bibitem{Durnin1987} J. J. Durnin, J. J. Miceli and J. H. Eberly, “Diffraction-free beams,” Phys.
Rev. Lett. 58, 1499–1501 (1987).
\end{thebibliography}

\end{document}